\documentclass[aps,pre,showpacs,twocolumn]{revtex4}
\usepackage{graphicx}
\usepackage{psfrag}
\usepackage{ae}
\usepackage{bm}
\usepackage[usenames]{color}
\newcommand{\rhot}{\rho_{\scriptscriptstyle\rm MAX}}
\newcommand{\tamanhofig}{8.6cm}

\begin{document}

\title{Monte Carlo simulations of $2d$ hard core lattice gases}

\author{Heitor C. Marques Fernandes}
\email{heitor@if.ufrgs.br}
\affiliation{Instituto de F{\'\i}sica, Universidade Federal do
Rio Grande do Sul \\ CP 15051, 91501-970 Porto Alegre RS, Brazil}

\author{Jeferson J. Arenzon}
\email{arenzon@if.ufrgs.br}
\affiliation{Instituto de F{\'\i}sica, Universidade Federal do
Rio Grande do Sul \\ CP 15051, 91501-970 Porto Alegre RS, Brazil}

\author{Yan Levin}
\email{levin@if.ufrgs.br}
\affiliation{Instituto de F{\'\i}sica, Universidade Federal do
Rio Grande do Sul \\ CP 15051, 91501-970 Porto Alegre RS, Brazil}

\begin{abstract}
Monte Carlo simulations are used to study
lattice gases of particles with extended hard cores on a
two dimensional square lattice. 
Exclusions of one and up to five nearest neighbors (NN) are considered.
These can be mapped onto
hard squares of varying side length,
$\lambda$ (in lattice units),
tilted by some angle with respect to the original
lattice. In agreement with earlier studies,
the 1NN exclusion
undergoes  a continuous order-disorder transition in the
Ising universality class.  Surprisingly, we find
that the lattice gas with exclusions of up to second nearest neighbors (2NN) 
also undergoes a continuous 
phase transition in the Ising universality class, while the
Landau-Lifshitz theory  predicts that this transition
should be in the universality class of the XY model with cubic anisotropy.  
The lattice gas of 3NN exclusions 
is found to undergo a discontinuous order-disorder transition,
in agreement with the earlier transfer matrix calculations and the 
Landau-Lifshitz theory.  
On the other hand,
the gas of 4NN exclusions once again exhibits a
continuous phase transition in the Ising universality class ---
contradicting the predictions of the Landau-Lifshitz theory.  
Finally,  the lattice gas of
5NN exclusions is found to undergo a 
discontinuous phase transition.

\end{abstract}

\maketitle
\date{\today}

\section{Introduction}
\label{section.introduction}

Since the early days of statistical mechanics lattice gases 
serve as the foundation on which many models of complex
physical systems are constructed.  These range from
simple fluids~\cite{Hill} to structural glasses and granular 
materials~\cite{RiSo03}. 
The simplest lattice gas consists of non-interacting 
particles which are constrained
to move on a lattice with a restriction that each lattice
site is occupied by at most one particle. This
model can be solved exactly, showing that the thermodynamics
is trivial and no phase transition is present.  
A lattice gas 
in which particles interact  
with their nearest neighbors can 
be mapped onto the Ising model in external field and  exhibits  
a first order  phase transition terminating in a critical point. 
The difficulty and the scarcity of exact
solutions has stimulated  
the development of various approximate theories aimed at treating
more complex interaction potentials. Examples of these are  
the high and the low temperature (or density) expansions~\cite{GaFi65}, 
generalization of Bethe's methods for Ising model (which became
known as cluster variational methods)~\cite{Burley72},
as well as the 
approximation schemes such as Rushebrook and Scoins~\cite{BeNi67}.
Since melting is
dominated by strong short ranged repulsive forces, 
a lattice gas 
in which particles interact exclusively 
through an extended hard core --- a particle on one site prevents
the neighboring sites from being occupied ---  
has attracted a particular attention~\cite{LaCu03jcp}. 
In these systems temperature plays 
no role since the interaction energy is 
infinite inside the exclusion region 
and vanishes outside. 
Surprisingly already with one
nearest neighbor exclusion (1NN), an order-disorder transition appears.  
The transition is purely entropic and is similar
to the freezing of hard spheres, except that 
now the transition is of second order.

In this paper we are interested 
in the phase transitions which occur in  hardcore lattice
gases as the size of the exclusion region is extended. 
We start with the 1NN  and progressively 
increase the exclusion region up to 
5 nearest neighbors (5NN).  To study the phase
transitions, grand-canonical Monte Carlo simulations are performed
with alternating attempts at inclusion/removal of particles followed
by some tentative diffusion moves. 
The grand-canonical ensemble was chosen because it allows for the
density fluctuations
which are particularly important since the
thermal fluctuations are absent in these systems. 


The specific question that we would like to address here is
how does the order-disorder transition
depends on the range of the exclusion region 
and whether the 
universality class and the order 
of the phase transition can be predicted based purely on 
symmetry considerations.

The paper is organized as follows: 
section~\ref{section.simulation} defines the
relevant order parameters characterizing the ordered phases. 
Section~\ref{section.results}
presents the results of the grand-canonical Monte Carlo simulations,
and section~\ref{section.conclusions} gives the final discussion and
the conclusions.

\section{Exclusion Models, sublattices and order parameters}
\label{section.simulation}

A lattice gas is composed of particles whose positions are 
restricted to coincide with the vertices of a given lattice,
and where each site can be  occupied by, at most, one
particle. When there is no further interaction between the 
particles, this system presents a trivial thermodynamics and no phase 
transition. Instead, here we consider equilibrium
lattice gases where the range of exclusion extends over the neighboring 
sites. Increasing the radius of the
exclusion shell is equivalent to taking smaller lattice
cell sizes as one goes to the continuum limit. These 
exclusion shells may be interpreted in terms of geometric
hard core particles. All
cases considered here are equivalent to parallel hard
squares of size $\lambda$, either tilted or not. 
Since hard core potentials allow either zero
or infinite energies, temperature is not a relevant
parameter, the systems thus being athermal. As a consequence,
the only relevant independent parameters are the
volume ($V$) and the chemical potential ($\mu$),
the latter determining the average number of particles inside the volume.
In the grand-canonical Monte Carlo 
simulation, three kinds of trial movements are allowed: 
displacement, insertion, and deletion of particles~\cite{Frenkel}.
The simulations are done
in two stages.  First various short runs are performed to
locate the critical region.   Then 
a single long run at the value of $\mu$ close
to the putative critical chemical potential $\mu_c$ is carried out. 
From the time series
data of the long run, 
using histogram reweighting techniques~\cite{FeSw88,NeBa99,Janke02}, 
the relevant thermodynamic observables 
are accurately evaluated throughout the critical region.
Another set of measurements at different values of $\mu$ is
also performed to check the reliability of the extrapolation.
The diffusion trial movements are performed explicitly to help
achieve higher densities and to
to avoid  entrapment of the system in 
low density metastable states.

 
The lattice gases with hard core exclusion 
present a transition from a low density fluid-like
phase, in which all sites have the same occupational probability, to a
large density crystal-like phase in which the system is no longer 
translationally invariant. Thus, the lattice may be
conveniently divided into sublattices which in the high density
phase are differently populated.
Below we describe the exclusion shells considered in our simulations, 
with the corresponding sublattices.  

For the system sizes
considered in our simulations, the density $\rho$ and  
the compressibility 
$\kappa=L^2(\left\langle \rho^2\right\rangle -\left\langle \rho\right\rangle^2)/\rho^2$, 
where $\left\langle\ldots\right\rangle$ is the ensemble average, 
are not very sensitive to the
phase transition.  To obtain a quantitative information
we, therefore, build order parameters which become non zero as a
system passes from a fluid to an ordered phase. When the
transition is continuous, the order parameter $q$ (to be defined below) 
obeys the usual finite size scaling relation
\begin{equation}
q=L^{-\beta/\nu} f[ (\mu-\mu_c)L^{1/\nu}],
\label{eq.qscaling}
\end{equation}
where $f$ is a scaling function
and 
$\mu_c$ is the critical value of the chemical potential in the
thermodynamical limit, $L \rightarrow \infty$.
The staggered susceptibility $\chi$ measures the fluctuations of $q$,
\begin{equation}
\chi=L^2\left( \left\langle q^2\right\rangle - \left\langle q\right\rangle^2\right),
\label{eq.susc}
\end{equation}
and has the scaling
\begin{equation}
\chi=L^{\gamma/\nu} g[ (\mu-\mu_c)L^{1/\nu}],
\label{eq.chiscaling}
\end{equation}
where $g$ is another scaling function. The
exponents $\beta$, $\gamma$ and $\nu$ are related with the critical
behavior of the order parameter, susceptibility and the correlation
length, respectively, and obey the hyperscaling
relation $2\beta/\nu+\gamma/\nu=d$. In the simulations below, one way
to obtain $\nu$ is
from the behavior of $\partial\ln q/\partial\mu$, whose scaling,
from eq.~\ref{eq.qscaling} follows
\begin{equation}
\frac{\partial\ln q}{\partial\mu} = L^{1/\nu} h[ (\mu-\mu_c)L^{1/\nu}]
\label{eq.derlog}
\end{equation}
where $h(x)=f'(x)/f(x)$. An analogous equation is obeyed if we
change $q$ by $\chi$. Alternatively, $\nu$ can also be obtained from
the behavior of the shifted location of the maxima of $\chi$ and $\kappa$
which approach the thermodynamic limit with  $L^{1/\nu}$ corrections. 
Once $\nu$ is known, $\beta$ and $\gamma$ are obtained
from eqs.~\ref{eq.qscaling} and \ref{eq.chiscaling}.




\subsection{Nearest neighbor exclusion (1NN)}

\begin{figure}[floatfix]
\psfrag{1}{\large $\bm 1$}
\psfrag{a}{\large\textcolor{white}{$\bm 1$}}
\psfrag{b}{\large\textcolor{white}{$\bm 2$}}
\psfrag{2}{\large $\bm 2$}
\psfrag{3}{\large $\bm 3$}
\psfrag{4}{\large $\bm 4$}
\includegraphics[width=4.0cm]{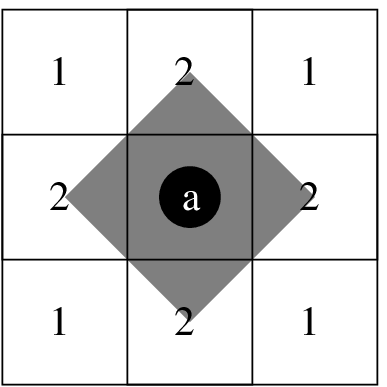}
\includegraphics[width=4.0cm]{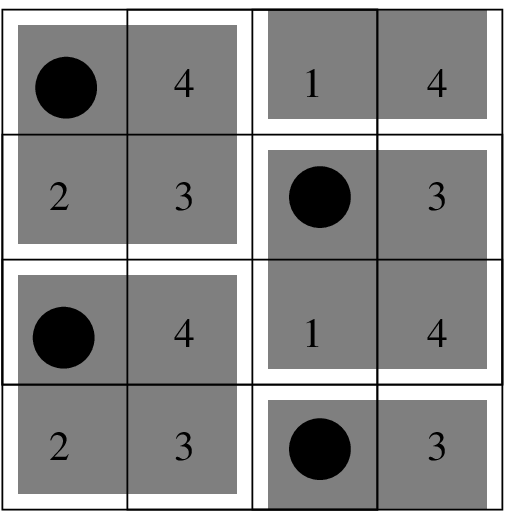}

(a) \hspace{3cm} (b)
\caption{Lattice division for the (a) 1NN and (b) 2NN exclusions:
two and four sublattices are needed, respectively. 
The 1NN case is equivalent to tilted, non overlapping 
hard squares 
of length $\lambda=\sqrt{2}$, while in the 2NN 
case the squares are not tilted and have  $\lambda=2$. 
The equivalent hard squares are shown in gray.}
\label{fig.sl1nn2nn}
\end{figure}

The simplest non trivial lattice gas consists of particles which  
preclude nearest neighbor sites from being occupied. 
As shown in Fig.~\ref{fig.sl1nn2nn}a,
the lattice may be subdivided into two sublattices, each site being 
surrounded by the sites of the other sublattice. This system, 
which can also interpreted either as $45^{\rm o}$ tilted hard-squares 
of linear size 
$\lambda = \sqrt{2}$ or as hard disks of 
radius $\sqrt{2}/2$, has been extensively
studied and here we present some results for the sake of
both completeness and comparison. 
Many different approaches have been used to describe its
properties on a square lattice: series 
expansions~\cite{Fisher63,GaFi65,BeNi67,BaEnTs80,BaFi94},
cluster variational and transfer matrix 
methods~\cite{Temperley62,Runnels65b,RuCo66,ReCh66,BeNi66,BeNi67,NiFa74a,WoGo80,KiSc81,PeSe88,GuBl02}, 
renormalization group~\cite{Racz80,HuCh91}, Monte Carlo 
simulations~\cite{BiLa80,Meirovitch83,HuMa89,Haas89,Yamagata95,HeBl96,HeBl98,LiEv00},
Bethe lattice~\cite{Burley61,BeNi67,Runnels67,Woodbury67,RoVa91,HaWe05a},
and more recently density functional theory~\cite{LaCu03pre}.
Moreover, this model has also been considered because of its
interesting mathematical~\cite{Runnels72,Baxter99,FeScEe05}
and dynamical~\cite{Murch81,Chaturvedi86,MeCaLoSc87,ErFrJa88,JaFrKn91,DiWaJe91,WaNiPr93a,EiBa98,Dickman01,HaWe05b,PoDi05}
properties.
These studies indicate that the system undergoes a
continuous phase transition belonging to the Ising universality class,
at $\mu_c\simeq 1.33$, when the concentration of particles is close to
$\rho=N/L^2\simeq 0.37$. Note that the maximum density is $1/2$.
In higher dimensions~\cite{Gaunt67,Cowley79,Racz80,HuMa90,Yamagata95,HeBl96,HeBlLu00,Panagiotopoulos05}
and other geometries (see~\cite{Runnels72} and references
therein), the same kind of transition is 
observed, also belonging to the Ising universality class.

The transition is from the disordered fluid phase, 
where both sublattices are equally occupied,
to a high density ordered phase with
most of the particles confined to one sublattice (see
Fig.~\ref{fig.sl1nn2nn}a).
The order parameter for this system is
\begin{equation}
q_1 = \frac{1}{\rhot}\left\langle | \rho_1 - \rho_2 |\right\rangle,
\label{eq.po1nn}
\end{equation}
where 
\begin{equation}
\rho_i = \frac{1}{L^2}\sum_{j\in {\cal C}_i} \sigma_j,\;\;\;i=1,2\;,
\label{eq.rhoi}
\end{equation} 
is the density in the $i$th sublattice ${\cal C}_i$ (the index $j$ running
over all sites in that sublattice),
$\rhot=1/2$ is the maximum equilibrium density. The  
variable $\sigma_j$ is $0$ if the site $j$ is empty and $1$ if it is occupied.
In the disordered low density regime 
$\rho_1=\rho_2$, and $q_1=0$ in the thermodynamic limit.
At higher densities, one sublattice becomes preferentially 
occupied, resulting in a finite value of the order parameter.
For finite systems the particles switch from one
sublattice to the other, analogous to the switching of the magnetization sign
in the Ising ferromagnet~\cite{LandauBinder}. The phase transition
appears only in the thermodynamic limit. 

\subsection{Nearest and next-nearest  neighbors exclusion (2NN)}

Extending the range of interaction, we now consider a lattice gas
such that each particle prevents both its nearest 
and next-nearest neighbors from being occupied. Thus, we define
four different sublattices labeled from 1 to 4 as 
shown in Fig.~\ref{fig.sl1nn2nn}b.
It should be noted 
that this system is equivalent to hard-squares with
linear size equal to two lattice spaces ($\lambda=2$), as is demonstrated in
Fig.~\ref{fig.sl1nn2nn}b; if a particle occupies a position on one sublattice,
say 1, the companion sites from sublattices 2 to 4 will be empty.
Thus, the maximum possible density, corresponding to a full lattice
occupation, is $\rhot=1/4$. 
Although this system has been considered many times in the 
last decades~\cite{BeNi66,Baker66,BeNi67,ReCh67,Mehta74,NiFa74c,DoScWaGr78,BiLa80,KiSc81,AkSh82,Slotte83,Huckaby83,AkSh84,AmKaGu84,RoVa91,WaNiPr93b,Uebing94,LaCu03jcp,EiBa98},
sometimes as the zero temperature limit 
of a soft-core potential~\cite{BiLa80,KiSc81,Slotte83,Uebing94}, there is
still uncertainty of its universality class.. 
The order, location, and even the existence 
of the transition depends on the approximation method used to study the
model. In table~\ref{tab.2nn} 
we show the density, the chemical potential, and the order of the 
transition obtained by several different methods.  Little is known about
its critical exponents, Ref.~\cite{KiSc81}.

In the low density regime, the system is fluid-like, 
without any long ranged order. As the density increases, there is
a transition from the disordered phase to a columnar 
phase~\cite{BeNi67,ReCh67,KiSc81,AkSh82,AkSh84,LaCu03jcp},
where half-filled columns (rows) intercalate with empty ones:
the system is ordered along one direction but fluid along the other.
Since there is no interaction between the particles (apart the
exclusion), there is no alignment between particles in the neighboring 
columns (rows), that is, there is no solid phase~\cite{ScLaCu03}.
As a consequence, the close packed configuration is not 
unique --- each column (or row,
but not both) may be shifted by one lattice space 
thus introducing an ${\cal O}(L)$ contribution
to the entropy at this maximum density. 
Interestingly, no transition
was found for the analogous
$d=3$ system of hard-cubes with $\lambda = 2$~\cite{Panagiotopoulos05} on a
cubic lattice. On the other
hand, nearest and next nearest neighbors exclusion in $d=3$
presents a weak
first order transition~\cite{Panagiotopoulos05}. Other lattices have
also been considered, see Refs.~\cite{Runnels72,BaEi84}
and references therein.

\begin{table}
\begin{center}
\begin{tabular}{|l|l|c|c|c|}
\hline\hline
$\mu_c$ & $\rho_c$ & Method & Order & Refs.\\
\hline\hline
4.7 & 0.24 & TM & cont (?) & \cite{KiSc81,AmKaGu84}\\
\hline
3.115 & 0.225 & CVM & & \cite{AkSh82} \\
\hline
3.889 & 0.222 & CVM & & \cite{AkSh84} \\
\hline
5.3 & 0.238 & TM & cont & \cite{ReCh67} \\
\hline
-- & -- & series & no & \cite{BeNi67,NiFa74c} \\
\hline
4 & 0.23 & TM & (?) & \cite{BeNi67} \\
\hline
-- & -- & TM & no & \cite{NiFa74c} \\
\hline
2.846 & 0.202 & Bethe & cont & \cite{BeNi67} \\
\hline
4.91 & -- & Interface & cont & \cite{Slotte83} \\
\hline
     &   & Bethe & first & \cite{RoVa91} \\
\hline
2.406 & 0.191 & DFT & cont  & \cite{LaCu03jcp} \\
\hline 
4.574 &  & Monte Carlo & cont & this work \\
\hline\hline
\end{tabular}
\caption{Chemical potential and the  density
at the order-disorder transition along with the technique
used and the putative order of the transition
for a lattice gas with nearest and next-nearest neighbors exclusion (2NN).
The symbol ``?'' is used when some uncertainty is acknowledged by the
authors. The several techniques used are: transfer matrix (TM),
cluster variational method (CVM), low and high series expansion, 
generalizations of the 
Bethe method, interface method~\cite{MuZi77}, density functional
theory (DFT) and Monte Carlo. Notice the
wide range of values found for the critical chemical potential.}
\label{tab.2nn}
\end{center}
\end{table}

The order parameter needed to study this transition 
is a generalization of the 1NN Eq.~\ref{eq.po1nn}:
\begin{equation}
q_2 = \frac{1}{\rhot} \left\langle |\rho_1 - \rho_3| + |\rho_2 - \rho_4|
\right\rangle .
\label{eq.po2nn}
\end{equation} 
In the fluid phase all the sublattices are equally occupied and $q_2=0$.
The order parameter becomes non zero when the symmetry 
between the sublattices is broken. For the columnar phase two 
sublattices are preferentially occupied, while 
the remaining two have a much smaller density. For example,
particles may be located on  sublattices 1 and 2 (as in
Fig.~\ref{fig.sl1nn2nn}b), while sublattices 3 and 4 are almost
empty etc.


\subsection{First, second and third neighbor exclusion (3NN)}

\begin{figure}[floatfix]
\begin{center}
\psfrag{1}{\large $\bm 1$}
\psfrag{2}{\large $\bm 2$}
\psfrag{3}{\large $\bm 3$}
\psfrag{4}{\large $\bm 4$}
\psfrag{5}{\large $\bm 5$}
\includegraphics[width=4.0cm]{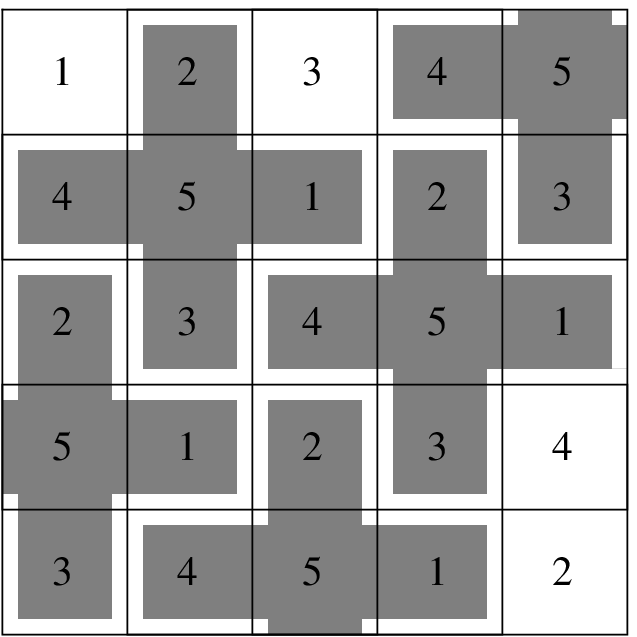}
\includegraphics[width=4.0cm]{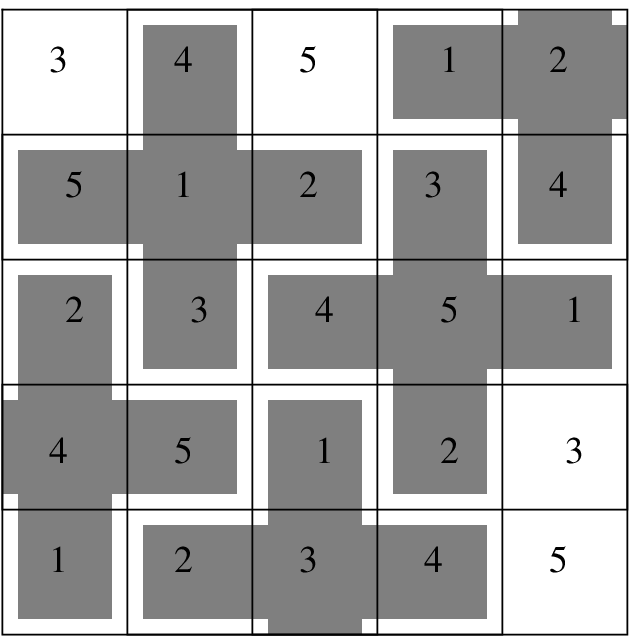}

(a) \hspace{3cm} (b)

\includegraphics[width=4.0cm]{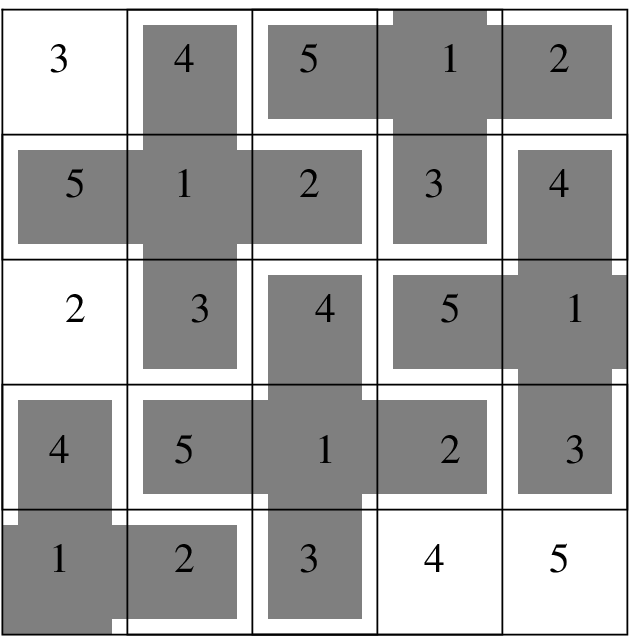}
\includegraphics[width=4.0cm]{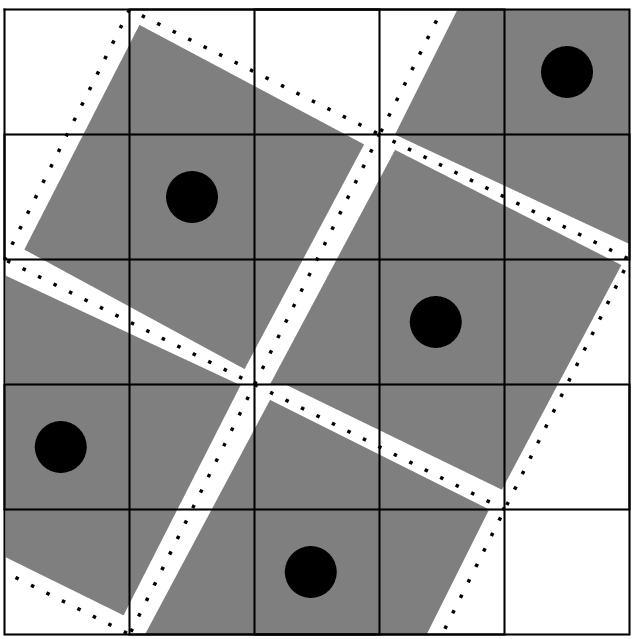}

(c) \hspace{3cm} (d)
\end{center}
\caption{Two possible labellings for the sublattices in
the 3NN case --- panels (a) and (b). 
For the same high density configuration in  
(a) all particles are on the same sublattice
(5 in this example), while in (b)
all sublattices are equally populated. For the labeling 
(b), panel (c) shows another configuration in which 
particles occupy only
one sublattice. Note that the 
ground states shown in panels (b) and (c) are chiral, in the sense
that the two  are the reflections of one another about the left-to-right
body diagonal.
The exclusion problem can be formulated either in terms of 
the symmetric cross-shaped pentamers
(shown in gray) or the tilted hard-squares of side length 
$\lambda = \sqrt{5}$
(d).}
\label{fig.sl3nn}
\end{figure}

The 3NN case has its range of exclusion extended up to
third nearest neighbors. As shown in Fig.~\ref{fig.sl3nn},
this can be interpreted both as hard-core pentamers~\cite{HePr74} or
slightly tilted squares with $\lambda=\sqrt{5}$.
Previous studies of this system were based on 
series expansions~\cite{BeNi67,OrBe82}, cluster variational and transfer 
matrix~\cite{BeOr66,BeNi66,BeNi67,NiFa74a,NiFa74b,OrBe82,EiBa05}, 
Bethe method~\cite{BeNi67}, {\it etc.}.
It is now known that at high densities, the system undergoes a 
first order transition to a doubly degenerated ordered 
phase~\cite{HaSt73},
the ground state configurations being related by chirality, 
see Figs. 2b and
2c. This is 
reflected in the two possible ways of introducing sublattices to describe 
the symmetry breaking: for each of the possible enantiomorph ground
states, there is a corresponding labeling (called A and B)
such that all particles belong to only one sublattice (see, for
example, Fig.~\ref{fig.sl3nn}a and b).

For a given configuration, since we do not know {\it a priori} which of the
two labellings is more relevant, we measure the following quantity for each
of them:
\begin{equation}
q_{\scriptscriptstyle\rm A} =
\sum_{i=1}^5 \sum_{j>i}^5 |\rho_i^{\scriptscriptstyle\rm A}
-\rho_j^{\scriptscriptstyle\rm A}|
\end{equation}
and equivalently for the labeling B. Note that in the above sums
the density of each sublattice appears four times.
 $\rho_i$ is given by eq.~\ref{eq.rhoi} for the chosen
labeling. This quantity is zero for both labellings
in the fluid phase,
where all sublattices are equally populated, while it becomes
non zero above the critical density (for one of the
labellings). The corresponding order parameter is the difference of
the two measures: 
\begin{equation}
q_3 = \frac{1}{4 \rhot} \left\langle|q_{\scriptscriptstyle\rm A} - 
q_{\scriptscriptstyle\rm B}|\right\rangle.
\label{eq.po3nn}
\end{equation}
Again, this order parameter is a natural extension of
$q_2$.
At the close packed configuration, where the density is $\rhot=1/5$, 
$q_{\scriptscriptstyle\rm A}=4\rhot$ and $q_{\scriptscriptstyle\rm B}=0$ (or vice-versa): 
the first labeling is such
that all particles sit on the same sublattice, while in the
second labeling all sublattices have the same density. In
both cases, however, $q_3=1$.

\subsection{4NN}

The subsequent situation considers exclusion of up to  four
nearest neighbor sites (4NN). As seen in Fig.~\ref{fig.5nn}a, this
is equivalent to tilted hard squares of side length 
$\lambda=2\sqrt{2}$.
Nisbet and Farquhar~\cite{NiFa74b}
studied this model through transfer matrices and concluded
that the transition is either continuous or very weak
first order. The close packed configuration shown in
Fig.~\ref{fig.5nn}a is similar
to the 1NN case, although less dense (the close 
packed density is $\rhot=1/8$), but it is not
unique: analogously to the 2NN case, the columns (which in this
case are tilted), are
independent and may slide along the diagonals. This does
not change, however, the sublattice that the particles occupy.  
Notice also
that this does not occur in the tilted diagonals of the
1NN and 3NN cases, the squares being always aligned.
Thus, the order
parameter should be insensitive to a diagonal displacement
of a whole (tilted) column, and it is enough to subdivide the lattice
in only two sublattices, in analogy with the 1NN case. 
Indeed, the transition to the ordered state may be analyzed with
the same order parameter, eq.~\ref{eq.po1nn}.


\subsection{5NN ($\lambda=3$)}

\begin{figure}[floatfix]
\begin{center}
\psfrag{1}{\large $\bm 1$}
\psfrag{2}{\large $\bm 2$}
\psfrag{3}{\large $\bm 3$}
\psfrag{4}{\large $\bm 4$}
\psfrag{5}{\large $\bm 5$}
\psfrag{6}{\large $\bm 6$}
\psfrag{7}{\large $\bm 7$}
\psfrag{8}{\large $\bm 8$}
\psfrag{9}{\large $\bm 9$}
\psfrag{a}{\large\textcolor{white}{$\bm 1$}}
\psfrag{c}{\large\textcolor{white}{$\bm 5$}}
\psfrag{d}{\large\textcolor{white}{$\bm 8$}}
\includegraphics[width=4.2cm]{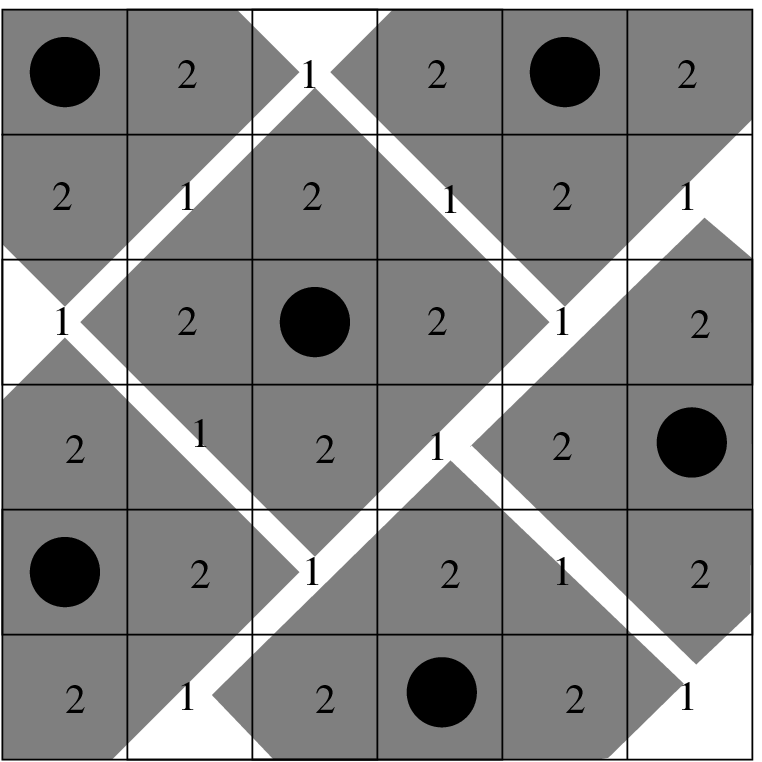}
\includegraphics[width=4.2cm]{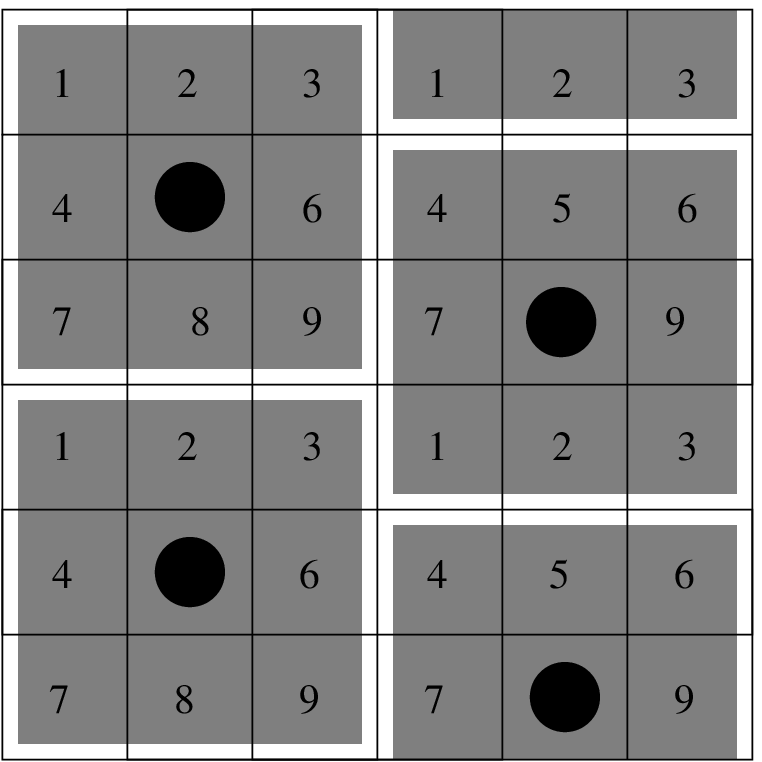}

(a) \hspace{3cm} (b)
\end{center}
\caption{Sublattice division for the exclusion of up to
(a) 4NN and (b) 5NN. These are
equivalent to hard squares with $\lambda = 2\sqrt{2}$
and 3, respectively. Notice that in both cases the ordered
phase is columnar (as in the 2NN case), but in the former,
the columns are along the diagonals.}
\label{fig.5nn}
\end{figure}

Finally, we considered a lattice gas
composed of hard-squares with linear size equal to
three lattice spaces, $\lambda=3$.
In terms of exclusion shells, this is also equivalent to
excluding up to the fifth nearest neighbor (5NN), as
shown in fig.~\ref{fig.5nn}b. The close packed configuration,
like in the 2NN ($\lambda = 2$) case, is not uniquely defined   
and has a density $\rhot=1/9$. In $d=3$, this
system undergoes a weak first order phase
transition as density increases~\cite{Panagiotopoulos05},
although there seems to be no information concerning the $d=2$
case. The corresponding
sublattice division is analogous to the 2NN case, only
that now nine sublattices are necessary,
the order parameter being
\begin{eqnarray}
q_5 &=& \frac{1}{2 \rhot}\left\langle
| \rho_1 - \rho_5 | + | \rho_1 - \rho_9 | + | \rho_5 - \rho_9 |
        \right. \nonumber \\
&&  + | \rho_2 - \rho_6 | + | \rho_2 - \rho_7 | + | \rho_6 - \rho_7 |\\
&&  + \left. | \rho_3 - \rho_4 | + | \rho_3 - \rho_8 | + | \rho_4 - \rho_8 |
\right\rangle 
\label{eq.pol3}
\end{eqnarray}
Eq.(\ref{eq.pol3}) uses the labeling  of sublattices depicted in 
Fig.~(\ref{fig.5nn}b).

\section{Results}
\label{section.results}


\subsection{1NN}

\begin{figure}[floatfix]
\psfrag{m}{\Large $\mu$}
\psfrag{r}{\Large $\rho$}
\psfrag{m1}{\Large $\mu$}
\psfrag{r1}{\Large $\kappa$}
\includegraphics[width=\tamanhofig]{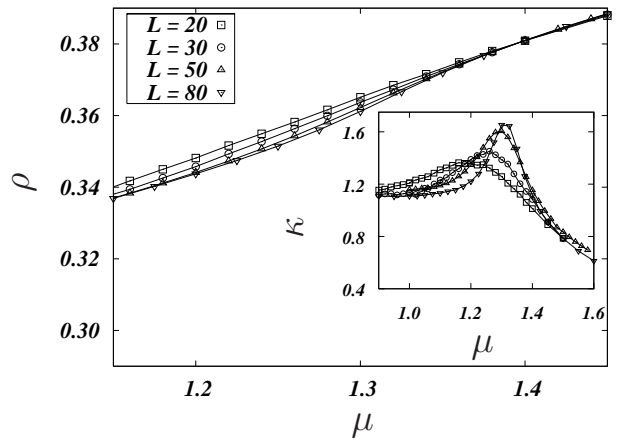}
\caption{Density $\rho$ as a function of the chemical potential
$\mu$ for the 1NN exclusion for several lattice sizes. The
inflection close to the transition point ($\mu_c\simeq 1.33$)
is very small, being noticeable only for larger system sizes.
Inset: compressibility $\kappa$ as a function of the chemical potential.}
\label{fig.rho1nn}
\end{figure}

\begin{figure}[floatfix]
\psfrag{x1}{\Large $\mu$}
\psfrag{y1}{\Large $q_1$}
\psfrag{x2}{\small\hspace{-1cm}$\left( \mu-\mu_c\right) L^{1/\nu}$}
\psfrag{y2}{\small$q_1L^{\beta/\nu}$}
\includegraphics[width=\tamanhofig]{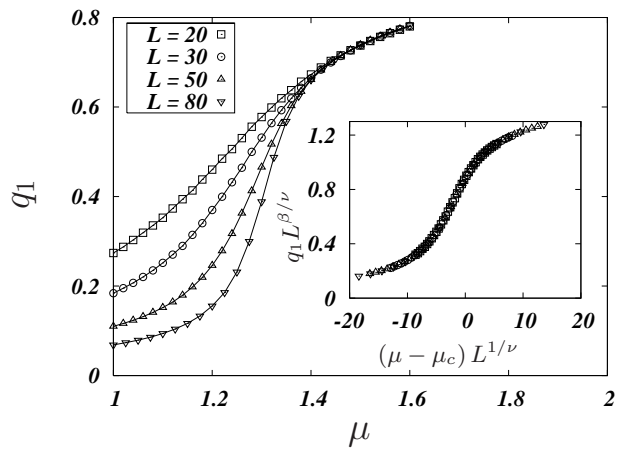}
\caption{Order parameter $q_1$, Eq.~\ref{eq.po1nn}, as a function of
the chemical potential $\mu$,
for the 1NN exclusion for several lattice sizes, along with
the corresponding data collapse (inset). 
The critical exponents used are those of the two
dimensional Ising model, $\nu=1$ and $\beta=1/8$.}
\label{fig.pocol1nn}
\end{figure}

In the case of nearest neighbors exclusion, the change in 
density at the transition is quite subtle, the inflection
point being only noticeable for the largest sizes simulated,
as seen in Fig.~\ref{fig.rho1nn}. The transition is more clearly
seen in the plot of the 1NN order parameter, Eq.~\ref{eq.po1nn}, as 
shown in Fig.~\ref{fig.pocol1nn} for increasing lattice sizes:
$q_1$ changes from a small value (fluid phase) to a value close
to unity (checkerboard pattern). The behavior of $q_1$ 
is characteristic of a second order phase transition and
indeed, all curves collapse onto a universal curve $f_1$, as 
shown in the inset of Fig.~\ref{fig.pocol1nn}, whose 
scaling obeys eq.~\ref{eq.qscaling}.
In accordance with the  
previous studies~\cite{GaFi65,ErFrJa88}, this transition 
occurs at $\mu_c\simeq 1.33$, corresponding to $\rho_c\simeq 0.37$, 
and belongs
to the universality class of the two dimensional Ising model,  with
exponents $\nu=1$ and $\beta=1/8$.

\begin{figure}[floatfix]
\psfrag{x2}{\small\hspace{-1cm}$\left( \mu-\mu_c\right) L^{1/\nu}$}
\psfrag{y2}{\small$\chi_1L^{-\gamma/\nu}$}
\psfrag{x1}{\Large $\mu$}
\psfrag{y1}{\Large $\chi_1$}
\includegraphics[width=\tamanhofig]{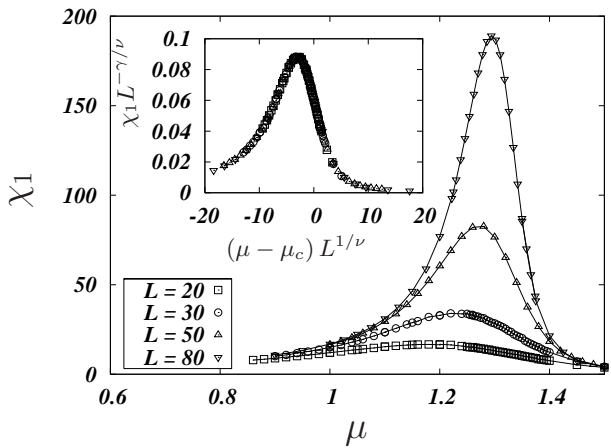}
\caption{Order parameter fluctuations $\chi_1$, Eq.~\ref{eq.susc},
 as a function of the chemical potential $\mu$ for the 1NN model using
several lattice sizes. In the inset, the corresponding data collapse is 
shown using the critical exponents of the bidimensional Ising model,
$\gamma=7/4$ and $\nu=1$.}
\label{fig.deltapocol1nn}
\end{figure}

Another quantity of interest, shown in Fig.~\ref{fig.deltapocol1nn},
is the staggered susceptibility $\chi_1 = L^2( \langle q_1^2 \rangle  -
\langle q_1\rangle^2 )$ which measures the order parameter fluctuations.  
As the system size increases, the susceptibility peak becomes 
higher and narrower, 
shifting to larger values of the chemical potential. All curves
can be collapsed onto a universal curve whose
scaling obeys eq.~\ref{eq.chiscaling},
as shown in the inset of Fig.~\ref{fig.deltapocol1nn}. Again,
the exponents are those of the two dimensional Ising model, 
$\gamma=7/4$ and $\nu=1$. 
Much less clear are the fluctuations in density, that is, the compressibility
$\kappa$, 
as shown in the inset of Fig.~\ref{fig.rho1nn}. Although the peak grows with
$L$, the sizes considered are still too 
small to try to obtain useful information
from the data collapse. Notice that in this case, the maximum is expected to
increase with the logarithm of $L$ ($\alpha=0$).

\subsection{2NN}

\begin{figure}[floatfix]
\psfrag{m}{\Large $\mu$}
\psfrag{r}{\Large $\rho$}
\psfrag{x}{\large $\mu$}
\psfrag{y}{\large $\kappa$}
\includegraphics[width=\tamanhofig]{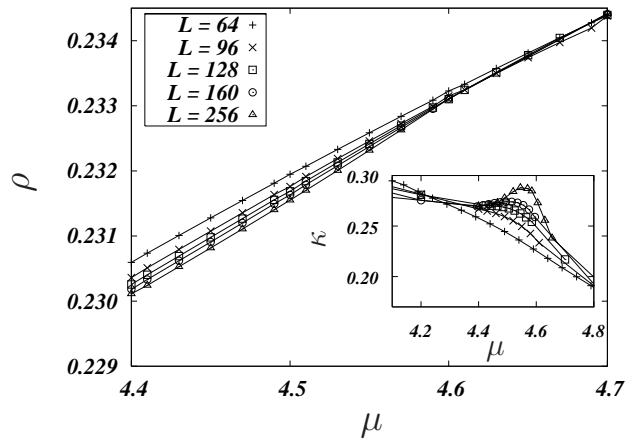}
\caption{Density as a function of chemical potential for several
lattice sizes for the gas of 2NN exclusion --- region where the 
inflection point appears (noticeable only for the larger system
sizes). Inset: compressibility $\kappa$ in the region where the
transition is found. Only for the largest lattice simulated the
maximum is prominent.}
\label{fig.rho2nn}
\end{figure}

\begin{figure}[floatfix]
\psfrag{x1}{\Large $\mu$}
\psfrag{y1}{\Large $q_2$}
\psfrag{x2}{\small\hspace{-1cm}$\left( \mu-\mu_c\right) L^{1/\nu}$}
\psfrag{y2}{\small$q_2L^{\beta/\nu}$}
\includegraphics[width=\tamanhofig]{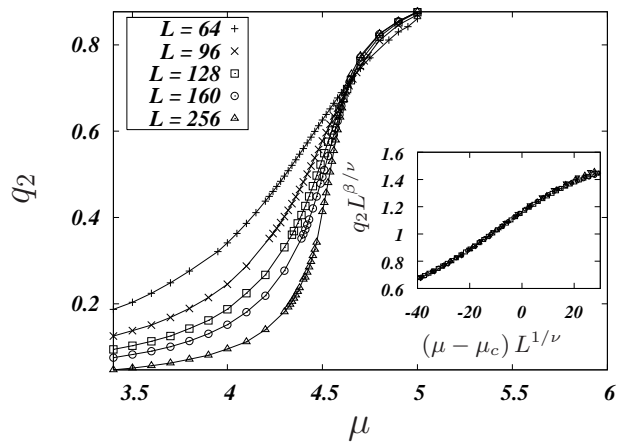}
\caption{Order parameter for 2NN  (eq.~\ref{eq.po2nn}) as a function
of the chemical potential for different lattice sizes. Small values of
$q_2$ signal that the system is disordered.
For $q_2$ closer to unity,
the system is in a columnar phase (see text). Inset: data collapse  
onto a universal curve for the 
larger system sizes $L=128, 160, 256, 320$ and 360. 
A very good collapse is obtained for $\mu_c=4.574$, $\nu=0.94$ 
and $\beta/\nu=0.125$, 
in close agreement with the exact Ising values. The collapse remains good
if we use the Ising value $\nu=1$.}
\label{fig.po2nn}
\end{figure}

Analogously to the previous case, when the range of exclusion 
increases to include the next nearest neighbors, it is very hard
to see the transition by only looking at the behavior of the
density as a function of the chemical potential. 
In the critical region only
for larger system sizes the inflection point becomes 
noticeable, as is shown in 
Fig.~\ref{fig.rho2nn}, and the change in density is roughly one 
order of magnitude smaller as compared with the 1NN case.
On the other hand, the order parameter $q_2$, Eq.~\ref{eq.po2nn}, shows a
fast increase between $\mu=4$ and 5, Fig.~\ref{fig.po2nn}, as the
system enters the ordered (columnar) phase from the disordered (fluid) one.
The the existence, location, order, and exponents of
this transition have remained elusive, as can be 
seen from a very broad range of approximate values presented 
in table~\ref{tab.2nn} and the lack of any estimates for the critical
exponents. We find that neither the density nor the order parameter 
$q_2$ present any signs of discontinuity.   
Within the range of lattice sizes considered no hysteresis,  
discontinuities,  nor any  evidences of the coexistence 
in the density and or the order parameter histograms has been observed.
We conclude that the transition is
continuous, in agreement with most of the previous studies. 

\begin{figure}[floatfix]
\psfrag{x}{\Large\hspace{-5mm} $1/L$}
\psfrag{y}{\Large\hspace{-8mm} $\mu_c-\mu_{\scriptscriptstyle\rm max}$}
\psfrag{x1}{\small $L$}
\psfrag{y1}{\small\hspace{-8mm} $\displaystyle\left.\frac{\partial\ln q_2}{\partial\mu}\right|_{\scriptstyle\rm max}$}
\includegraphics[width=\tamanhofig]{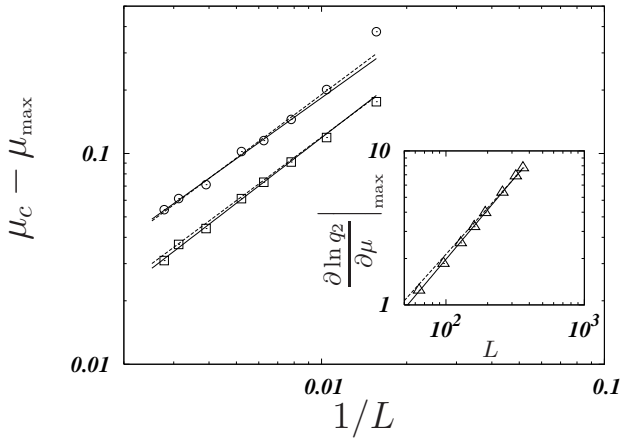}
\caption{Several estimates of the exponent $\nu$ (2NN): the position of
the maxima of eq.~\ref{eq.derlog} (top curve) 
and the susceptibility $\chi_2$ (bottom),
that shift as $L^{-1/\nu}$, as a function of $L$. The solid lines are 
power-law fits neglecting the smaller sizes, from which we get
$\nu\simeq 1.042$ and $\nu\simeq 0.968$, top and bottom, respectively.
These values are very close to the Ising $\nu=1$, as can
be seen by the fits fixing the exponent to this value (dashed lines).
Also from the fit of the location of the maximum of the susceptibility we
have an independent estimate of the transition: $\mu_c\simeq 4.578$.
Inset: the height of the maximum of  eq.~\ref{eq.derlog}, increasing
as  $L^{1/\nu}$, as a function of $L$. Again the dashed line is a
fit with $\nu=1$, while the solid line is the best fit with $\nu\simeq 0.937$.}
\label{fig.nu2nn}
\end{figure}

Indeed, the $q_2$ data can be collapsed onto a universal curve 
(as in Eq.~\ref{eq.qscaling}), with $\mu_c\simeq 4.574$, $\beta/\nu=0.125$
and $\nu=0.94$. Remarkably, $\beta/\nu$ has the same value of the 
(exact) Ising model, $1/8=0.125$. Although $\nu$ is a little bit smaller than
the Ising value, a still good collapse is obtained with $\nu=1$ and,
as we will see below, is in agreement with other measures of this
exponent. The density, measured for several systems sizes at 
$\mu_c$, or at the peak of the susceptibility extrapolate
when $L\to\infty$ to $\rho_c=0.233$. 
The value of $\mu_c$ is also compatible with the
crossing point of the Binder cumulant for the order parameter and with the extrapolated
position of the susceptibility, fig.~\ref{fig.nu2nn}, $\mu_c\simeq 4.578$.

\begin{figure}[floatfix]
\psfrag{x2}{\small\hspace{-7mm}$\left( \mu-\mu_c\right) L^{1/\nu}$}
\psfrag{y2}{\small\hspace{-2mm}$\chi_2L^{-\gamma/\nu}$}
\psfrag{x}{\Large $\mu$}
\psfrag{y}{\Large $\chi_2$}
\psfrag{x1}{\small $L$}
\psfrag{y1}{\small $\chi_2^{\scriptscriptstyle\rm max}$}
\includegraphics[width=\tamanhofig]{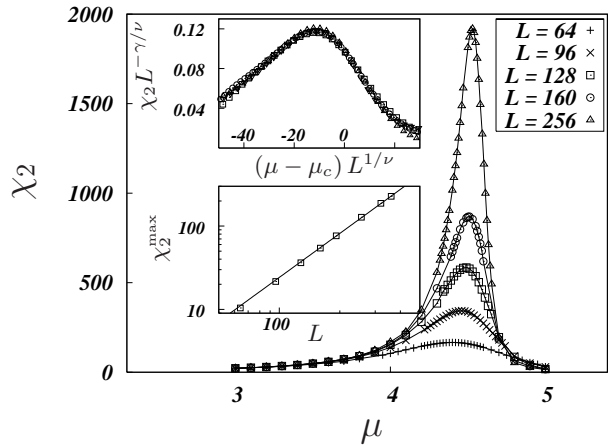}
\caption{The staggered susceptibility $\chi_2$ 
as a function of $\mu$ for several lattice sizes in the 2NN case.
Inset: data collapse (top) onto a universal curve with 
exponent $\gamma$ obtained from fitting the height of the maximum of $\chi_2$
as a function of $L$ (bottom), $\gamma/\nu=1.755$, along with $\nu=1$.}
\label{fig.deltapo2nn}
\end{figure}

We also measured the order parameter and the density fluctuations --- that is, 
the staggered susceptibility $\chi_2$, eq.~\ref{eq.susc}, and the compressibility 
$\kappa_2$, respectively, as shown in Figs.~\ref{fig.deltapo2nn} and 
\ref{fig.rho2nn} (inset) as a function of the chemical potential. In both 
cases, as the system size increases the curves get  less broad and their 
height becomes larger. However, the maximum of $\kappa_2$ only becomes
noticeable for the larger simulated lattices and no reliable 
information can be
obtained from its scaling, within the range of $L$ considered here. 
The shift of the position of the peak of $\chi_2$ from $\mu_c$ behaves as $L^{-1/\nu}$
(in the same way as the position of the maximum in eq.~\ref{eq.derlog}), from where 
$\nu$ can be estimated, as shown in fig.~\ref{fig.nu2nn}. In addition, the
height of the peak of eq.~\ref{eq.derlog} increases with $L^{1/\nu}$.
The exponent $\nu$ evaluated from both measurements is very close to 
the Ising value, see caption of fig.~ \ref{fig.nu2nn} for
details. We conjecture, therefore, that the exact 
value is indeed $\nu=1$. 
The height of the peak of $\chi_2$ increases as $L^{\gamma/\nu}$, and the
fit yields $\gamma/\nu=1.755$ (bottom inset of fig.~\ref{fig.deltapo2nn}), 
giving an excellent data collapse 
as can be seen in the top inset of fig.~\ref{fig.deltapo2nn}. Again,
this value is very close to the known exact value
for the Ising model, $\gamma=7/4=1.75$. 
The fact that the lattice gas of 2NN exclusion is 
in the same universality class as the Ising model is quite remarkable
considering that the symmetries of the ground state and of the order 
parameter are so very
different from those of 1NN model.

\subsection{3NN}

\begin{figure}[floatfix]
\psfrag{x}{\LARGE $\mu$}
\psfrag{y}{\LARGE $\rho$}
\psfrag{x1}{\large $\mu$}
\psfrag{y1}{\large $q_3$}
\includegraphics[width=\tamanhofig]{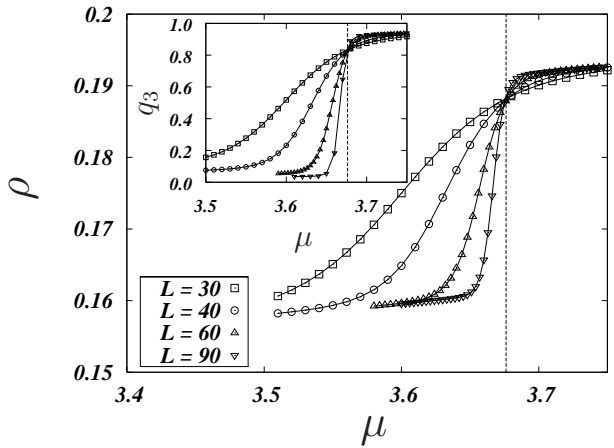}
\caption{Density as a function of chemical potential for the 3NN
exclusion model. At low densities the system behaves as a fluid
and undergoes a first order phase transition to an ordered phase
as the density increases. The vertical dashed line at $\mu=3.6762$ is
the infinite size extrapolation obtained with matrix methods~\cite{EiBa05}
and agrees  well with the crossing point of the curves. Inset: 
the order parameter as a function of the chemical potential.}
\label{fig.rho3nn}
\end{figure}

\begin{figure}[floatfix]
\psfrag{x}{\Large $\mu$}
\psfrag{y}{\Large $\kappa$}
\psfrag{x1}{\hspace{-8mm} $(\mu-\mu_c)L^2$}
\psfrag{y1}{\hspace{-4mm} $\kappa L^{-2}$}
\includegraphics[width=\tamanhofig]{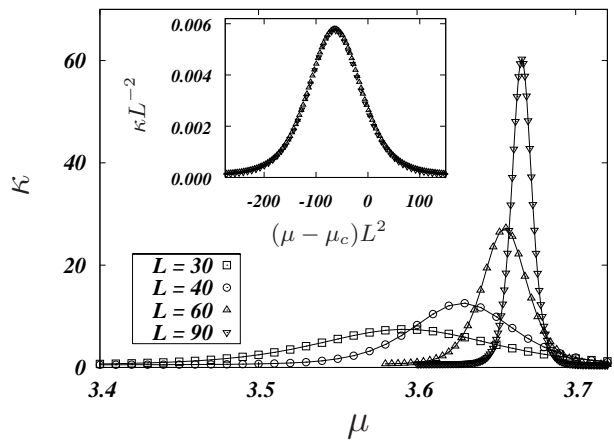}
\caption{Compressibility $\kappa$ as a function of the chemical 
potential for the 3NN exclusion for several lattice sizes. 
Inset: after rescaling the height and the width of the curves by $L^2$ and
$L^{-2}$, respectively, an excellent collapse is observed for the two
largest systems with $\mu_c=3.6746$.}
\label{fig.comp3nn}
\end{figure}

\begin{figure}[floatfix]
\psfrag{x}{\Large $\mu$}
\psfrag{y}{\Large $\chi_3$}
\psfrag{x1}{\hspace{-8mm} $(\mu-\mu_c)L^2$}
\psfrag{y1}{\hspace{-4mm} $\chi_5 L^{-2}$}
\includegraphics[width=\tamanhofig]{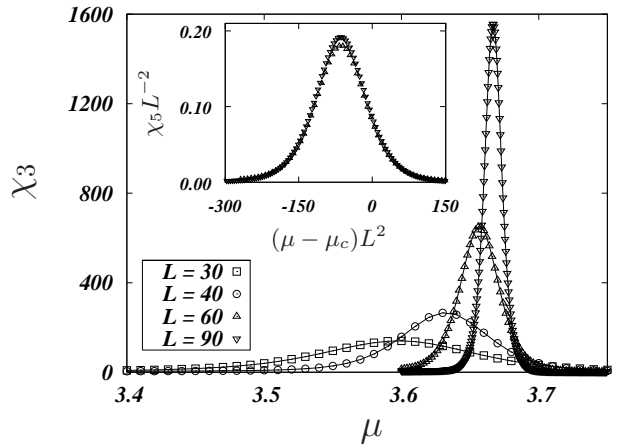}
\caption{Staggered susceptibility $\chi_3$ as a function of the chemical 
potential for the 3NN exclusion case for several lattice sizes. 
Inset: after rescaling the height and the width of the curves by $L^2$ and
$L^{-2}$, respectively, an excellent collapse is observed with
$\mu_c=3.6746$.}
\label{fig.susc3nn}
\end{figure}

The lattice gas with exclusion of up to three nearest neighbors (3NN) 
is important since it presents,
differently from the previous cases, a first order transition,
showing that the nature of the transition depends on the range 
of exclusion, as was observed previously. Recently,  Eisenberg and 
Baram~\cite{EiBa05} precisely located the transition, using 
matrix methods, obtaining $\mu_c\simeq 3.6762$. 
In Fig.~\ref{fig.rho3nn} and its inset, 
both the density and the order parameter, respectively, show a rapid 
rise as the chemical potential increases, passing from 
a fluid, low density phase, to an ordered, high density phase. Notice
also the good agreement between the matrix prediction for the transition and
our Monte Carlo data.
The first order character of this transition
is clearly seen in the behavior of the compressibility~\cite{OrBe82} 
and the fluctuations of the order parameter, shown in 
Figs.~\ref{fig.comp3nn} and \ref{fig.susc3nn}. All the
relevant scaling is with the volume of the system $L^2$ ($\nu=1/d$~\cite{FiNi82}), as can be seen
in the inset of both figures. Excellent data collapses were 
obtained with $\mu_c=3.6746$, very close to the value 
calculated using the transfer matrix~\cite{EiBa05}. 
This value is also very close to the ones 
obtained by extrapolating the positions
of the compressibility and the susceptibility peaks as a function of the
system size, and with the crossing points of the density and the
order parameter for the two largest system sizes.

\subsection{4NN}

When excluding up to four neighbors, the close packed
state resembles both the 1NN and 2NN case. On one hand,
particles are located on one of the two sublattices, arranged
along the diagonals of the system (although skipping
one every two sites); on the other hand, these diagonals
are independent ``columns'' that may slide, originating
a $2^L$-degenerated ground state with density $1/8$. Surprisingly,
within the size range available, the transition appears to be
continuous and in the Ising universality class, contradicting 
the Landau-Lifshitz theory which predicts that this transition should
be of first order~\cite{DoScWaGr78}.

\begin{figure}[floatfix]
\psfrag{x}{\Large\hspace{-5mm} $1/L$}
\psfrag{y}{\Large\hspace{-8mm} $\mu_c-\mu_{\scriptscriptstyle\rm max}$}
\psfrag{x1}{\small $L$}
\psfrag{y1}{\small\hspace{-10mm} $\displaystyle\left.\frac{\partial\ln q_1}{\partial\mu}\right|_{\scriptstyle\rm max}$}
\includegraphics[width=\tamanhofig]{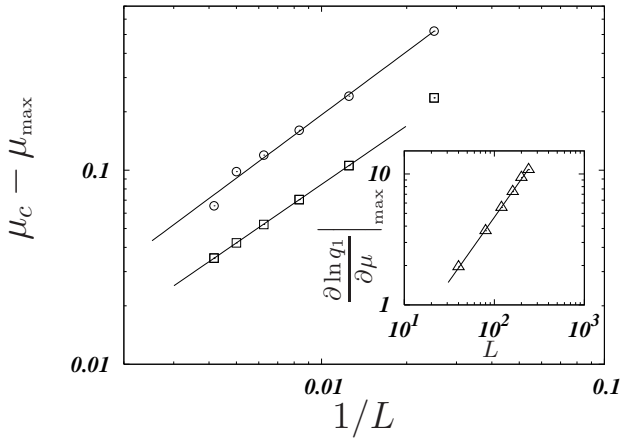}
\caption{Several estimates of the exponent $\nu$ (4NN): the position of
the maxima of eq.~\ref{eq.derlog} and the susceptibility $\chi_1$ (bottom),
that shift as $L^{-1/\nu}$, as a function of $L$. The solid lines are 
power-law fits neglecting the smaller sizes, from which we get
$\nu\simeq 0.923$ and $\nu\simeq 0.999$, top and bottom curves, respectively.
These values are indeed very close to the Ising one, $\nu=1$ (notice
also that the peaks of eq.~\ref{eq.derlog} are broader than those of
$\chi_1$, and their positions are less reliable).
Also from the fit of the location of the maximum of the susceptibility we
have an independent estimate of the transition point: $\mu_c\simeq 4.705$.
Inset: the height of the maximum of  eq.~\ref{eq.derlog}, increasing
as  $L^{1/\nu}$, as a function of $L$. The solid line is a fit with 
$\nu\simeq 1.029$.}
\label{fig.nu4nn}
\end{figure}

\begin{figure}[floatfix]
\psfrag{x1}{\Large $\mu$}
\psfrag{y1}{\Large $q_1$}
\psfrag{x2}{\small\hspace{-1cm}$\left( \mu-\mu_c\right) L^{1/\nu}$}
\psfrag{y2}{\small$q_1L^{\beta/\nu}$}
\includegraphics[width=\tamanhofig]{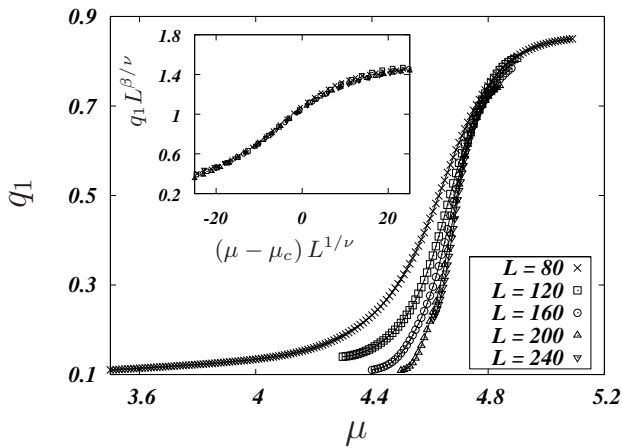}
\caption{Order parameter for the 4NN exclusion 
(eq.~\ref{eq.po1nn}) as a function
of the chemical potential for different lattice sizes. For large values of 
$q_1$, the system is in a columnar phase with columns aligned  
along the diagonals. Inset: collapse of data
onto a universal curve with $\nu=1$, $\beta=0.125$ and
$\mu_c=4.705$.  Equally good collapse is obtained with
a broad range of  $\beta$.}
\label{fig.po4nn}
\end{figure}

As in the 2NN case, several estimates of $\nu$ can be obtained.
From the positions of the maxima of eq.~\ref{eq.derlog} and $\chi_1$
we obtain, fig.~\ref{fig.nu4nn}, $\nu\simeq 0.923$ and $\nu\simeq 0.999$,
respectively.
The last one is more reliable since the peaks of $\chi_1$ are less
broad, thus allowing a more precise location of the maxima. A further
measure of $\nu$ is obtained from the height of eq.~\ref{eq.derlog}
and gives $\nu\simeq 1.029$. As in the 2NN case, these values
oscillate around the Ising  $\nu=1$ and, again, we conjecture
that this is the exact value. Indeed, this exponent also
fits quite well the data. As before, the transition
is not clear from the behavior of the density, 
while a properly defined order parameter is able to capture it 
well, fig.~\ref{fig.po4nn}. The
transition from the ordered, tilted columnar phase, to the fluid phase
is found to occur at $\mu_c\simeq 4.705$, corresponding
to $\rho_c\simeq 0.110$.

The order parameter and its collapse are shown
in fig.~\ref{fig.po4nn}. The collapse is not very
sensitive to the chosen value of $\beta$, and any
value in the broad range between 0.12 and 0.14 will
do well. We thus present the collapse with $\beta/\nu=0.125$,
the Ising value. Although this is rather arbitrary,
the closeness of both $\nu$ and $\gamma$ (see below)
to the Ising values inspire us to believe that indeed
this system is in the Ising universality class.

\begin{figure}[floatfix]
\psfrag{x2}{\small\hspace{-7mm}$\left( \mu-\mu_c\right) L^{1/\nu}$}
\psfrag{y2}{\small\hspace{-2mm}$\chi_1L^{-\gamma/\nu}$}
\psfrag{x}{\Large $\mu$}
\psfrag{y}{\Large $\chi_1$}
\psfrag{x1}{\small $L$}
\psfrag{y1}{\small $\chi_1^{\scriptscriptstyle\rm max}$}
\includegraphics[width=\tamanhofig]{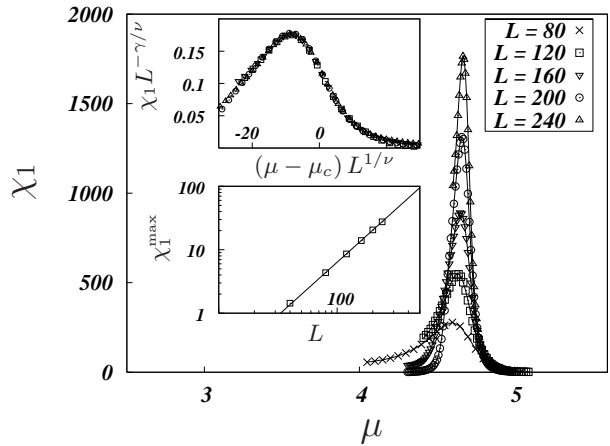}
\caption{The staggered susceptibility $\chi_1$ 
as a function of $\mu$ for several lattice sizes for 4NN exclusion.
Insets: (top) data collapse onto a universal curve with 
exponent $\gamma$ obtained from fitting the height of the maximum of $\chi_1$
as a function of $L$ shown in the bottom inset; 
$\gamma/\nu=1.681$ and  $\nu=1$.}
\label{fig.deltapo4nn}
\end{figure}

The fluctuations of the order parameter are shown in 
fig.~\ref{fig.deltapo4nn}. The dependence of the
susceptibility peak on the system size is
shown in the inset (below), yielding 
$\gamma/\nu=1.681$.  With these values
for the exponents, all curves can be collapse onto a
universal one. An equally good collapse is obtained
if instead of this value of $\gamma$ we use the
Ising one, 1.75.  We conclude that the phase transition
in the lattice gas with 4NN exclusion is in the Ising universality
class.

\subsection{ 5NN ($\lambda=3$)}

The lattice gas with exclusion of up to five nearest neighbors is
equivalent to hard square particles with side length $\lambda=3$.
In this case, as 
for the 2NN and 4NN, the system does not have a uniquely
defined close packed configuration which can bring
problems when approximated (analytical) methods such
as series expansions and cluster variational methods
are applied.  

\begin{figure}[floatfix]
\psfrag{x}{\LARGE $\mu$}
\psfrag{y}{\LARGE $\rho$}
\psfrag{q5}{\large $q_5$}
\psfrag{m}{\large $\mu$}
\includegraphics[width=\tamanhofig]{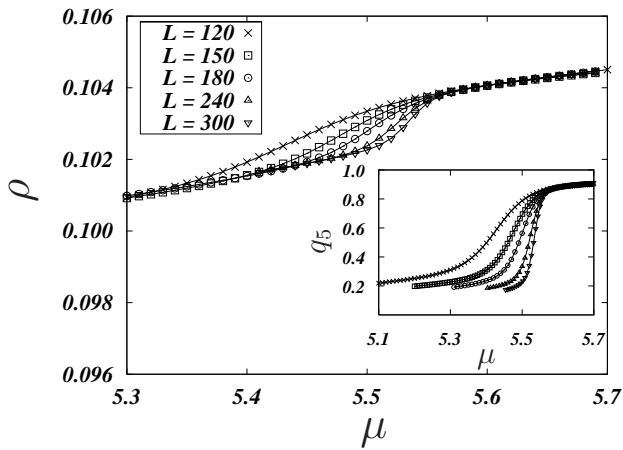}
\caption{Density as a function of the chemical potential for 
5NN exclusion ($\lambda=3$) for different lattice sizes.
The steepness increases as the system develops a
discontinuity. Inset: order parameter $q_5$ as a function of
$\mu$ for the same system sizes.
}
\label{fig.rhol3}
\end{figure}

\begin{figure}[floatfix]
\psfrag{x}{\Large $\mu$}
\psfrag{y}{\Large $\chi_5$}
\psfrag{x1}{\hspace{-8mm} $(\mu-\mu_c)L^2$}
\psfrag{y1}{\hspace{-4mm} $\chi_5 L^{-2}$}
\includegraphics[width=\tamanhofig]{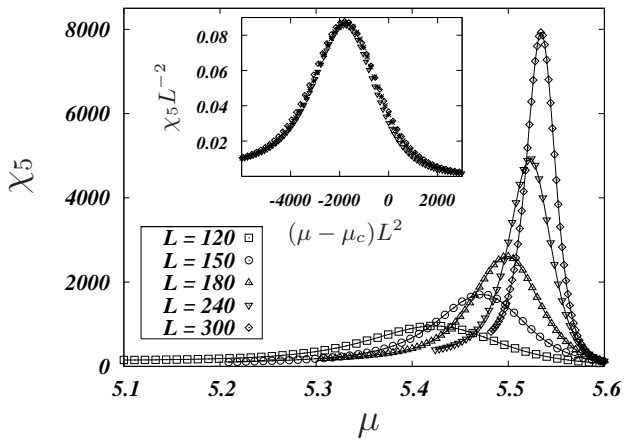}
\caption{Staggered susceptibility $\chi_5$ as a function of the chemical 
potential for the 5NN  exclusion ($\lambda=3$) for several lattice sizes. 
Inset: after rescaling the height and the width of the curves by $L^2$ and
$L^{-2}$, respectively, a good collapse is observed for the larger simulated 
sizes with $\mu_c=5.554$}
\label{fig.chi5}
\end{figure}

When the density is measured as a function of 
the chemical potential, fig.~\ref{fig.rhol3}, there is a
fast increase between $\mu=5.5$ and 5.6, and the
curves become steeper as the system size gets larger, signaling the
ordering of the system. However, the increase in the 
density is rather small, while 
a much more pronounced change is observed for the order parameter,
eq.~\ref{eq.pol3}.
The fluctuations of the order parameter, shown in Fig.~\ref{fig.chi5},
can be well collapsed with a $L^{-2}$ scaling (the
volume of the system), at least for the larger sizes considered.  This is a
signature of a first order transition. 
The location of the maximum depends on 
$L$ and obeys the $L^{-2}$ shift with  $\mu_c \simeq 5.554$.
There seem to be, however, strong finite size effects for smaller
system sizes, as observed when trying to collapse all the curves --- only 
for the larger sizes considered the collapse is sufficiently good. 
For the density fluctuations (not shown), even stronger effects are seen.
This is probably
due to the fact that the increase in the compressibility with $L$
is very small and the non singular part of the free energy  
can not be discarded when
compared to the singular one. This seems to be true for all the cases
considered in this paper (being least pronounced for 3NN exclusion).


\section{Conclusions}
\label{section.conclusions}

We used the grand-canonical Monte Carlo simulations to study the
order-disorder transition in 
lattice gases composed of particles with extended hard core on a
two dimensional square lattice. 
In agreement with earlier studies~\cite{GaFi65},
the 1NN exclusion --- corresponding 
to hard squares of length  $\sqrt{2}$ tilted by $45^{\rm o}$ --- 
was found to undergo  a continuous order-disorder transition in the
Ising universality class.  
The lattice gas with exclusions of up to second nearest neighbors (2NN) 
was also found to undergo a continuous  
phase transition in the Ising universality class, while the
Landau-Lifshitz (LL) theory~\cite{DoScWaGr78} predicts that this transition should be
in the universality class of the XY model with cubic anisotropy.  
The lattice gas with exclusions
of up to 3NN was found to undergo a discontinuous order-disorder transition,
in agreement with the earlier transfer matrix calculations~\cite{EiBa05} and the 
LL theory~\cite{DoScWaGr78}.  
On the other hand,
the gas of 4NN exclusions once again exhibited a
continuous phase transition in the Ising universality class --- a conclusion
which contradicts the LL prediction of a first order phase transition~\cite{DoScWaGr78}.  
Finally,  the lattice gas with
exclusions of up to 5NN was found to undergo a 
discontinuous phase transition, in agreement with the LL theory.

The failure of the symmetry based Landau-Lifshitz theory to
correctly predict the universality class and the order of the phase 
transition for some two dimensional lattice gases is not very surprising.  
It has been known for a long
time that while the Landau-Lifshitz
theory works well in three dimensions, its foundations are
much  shakier in 2d.  For example, one of its requirements
for the existence of a continuous 
phase transition is the absence of third order invariants 
of the irreducible 
representation of the space group of the high symmetry phase 
to which the order parameter belongs.  In 2d this
condition is known to break down --- both three and four state
Potts models undergo a continuous phase transition, even though
their Landau-Ginzburg-Wilson (LGW) Hamiltonians 
possesses third order invariants~\cite{Baxter73,MiSt74}.  In view of this problem,  
Domany et al. have avoided the use of this ``Landau rule'' 
in their classification of the continuous order-disorder transitions of 2d
lattice gases~\cite{DoScWaGr78}.  Nevertheless, the fact that one of the Landau
rules fails, can already serve as a strong warning against a complete
reliance on the symmetry arguments for two dimensional systems. 
The fundamental difficulty is that 2d 
is the lower critical dimension for lattice gases.  The fluctuation
are very important and can easily 
modify the order 
of the phase transitions from the one predicted by mean-field.
Furthermore, in 2d one can no longer rely on 
the renormalization group theory (RG) to justify a simple 
low order polynomial form of the LGW Hamiltonian. 
Unlike in 3d, where it is possible to show that the  
high order invariants in the LGW Hamiltonian  
renormalize to zero under the action of
RG, in 2d all the operators
are equally relevant --- thus invalidating a simple low order
polynomial form of the LGW Hamiltonian. In the case of 
the isotropic XY model
the situation is particularly dire since  
the symmetry based construction of the LGW Hamiltonian 
fails to consider the topological excitations which lead to the 
Kosterlitz-Thouless phase transition in this 
system~\cite{KoTh73,Be71a,Be71b}. It is, therefore,
unlikely that symmetry considerations alone will be
sufficient to predict the universality class and the transition order 
for all 2d lattice gases. 

Nevertheless, while the second order transition for the gas of 4NN exclusions
is completely incompatible with even the modified version of the
LL theory (in which the third order invariants are allowed), 
the case of 2NN is not nearly as bad.  The LGW Hamiltonian 
for the lattice gas with 2NN exclusion~\cite{DoScWaGr78} is 
\begin{eqnarray}
H=\int {\rm d}^2 {\bf x} &\{&(\nabla \psi_1)^2+(\nabla \psi_2)^2
+r(\psi_1^2+\psi_2^2) + \nonumber\\
&&u(\psi_1^2+\psi_2^2)^2+ v(\psi_1^4+\psi_2^4)\} \,
\label{lgw}
\end{eqnarray}
where $\psi_i$ with $i=1,2$ are the two components of the order
parameter belonging to the two dimensional irreducible representation of 
the space group $P4mm$.

In the absence of cubic anisotropy (last term of eq.~(\ref{lgw})) this
Hamiltonian  will not have a 
phase transition since the massless Goldstone excitations 
will destroy the long-range order at any finite temperature. 
Furthermore,  eq.~(\ref{lgw}) does not account for the topological defects
(vortices) which are responsible for the phase transition in the real XY
model~\cite{KoTh73,Be71a,Be71b} 
and the appearance of a pseudo-long-range 
order.  Thus, in the passage 
from a microscopic Hamiltonian to the coarse
grained LGW description, important physics has been left 
behind~\cite{LeDa90,Le91}.  
Clearly under such conditions no conclusions based on this LGW Hamiltonian
can be trusted.  Nevertheless, if we take the form of eq. (\ref{lgw}) more
seriously than perhaps it deserves, the Ising criticality is
not incompatible with its structure.  Based on the
general properties of RG flows, we expect that eq. (\ref{lgw}) 
will have an Ising fixed point 
at which the isotropic fourth order invariant will vanish.  
At this fixed
point, the LGW Hamiltonian will decouple into two Ising-like
Hamiltonians for each component of the order parameter.  
The symmetry arguments, however, cannot tell us 
if the Ising fixed point is stable, 
or even if it is accessible to the renormalization group flow. 
Nevertheless, unlike the second order phase transition 
for 4NN exclusions,
the Ising criticality for 2NN
exclusions does not, in principle, invalidate 
the predictions of the LL theory.
This not withstanding, it is quite surprising that in a multidimensional
parameter space with an infinite number of 
fixed points (fixed lines) known to
exist for the real XY model with cubic anisotropy~\cite{JoKaKiNe77}, the 2NN 
model just happens to be in the region dominated by the Ising fixed
point.


For lattice gas with 4NN exclusion the simulations find 
a continuous order-disorder transition in the Ising universality class.
We note, however, that within a simulation it is impossible to 
be absolutely certain that the transition is  
not a very weakly first order as happens, for example,  
for the five states Potts model.  
There is, however,  a big difference between  
$q$ states Potts models and hard core lattice gases studied in this
paper.
For Potts model, 
the mean-field theory~\cite{MiSt74} predicts that all transitions
for $q\ge 3$ are of first order.  
The fluctuations, however, 
modify the order of $q=3$ and $4$ to second, 
and turn  $q=5$ very weakly first order~\cite{Baxter73}. 
Importance of fluctuations decreases for larger values
of $q$ and the mean-field becomes progressively more 
accurate --- the phase transitions for $q>4$ are all of first order.
For lattice gases, on the other hand,
mean-field theory (LL) 
predicts the transition to be of second order
up to 2NN exclusions and first order afterwards.  
Indeed, we find continuous phase transitions for
1NN and 2NN exclusions and a first order transition for 3NN. 
Unlike the case of Potts model, however, the agreement with the 
mean-field  theory is
broken precisely at 4NN and re-established again for 5NN. 
There are no precursor fluctuation induced second order phase
transitions (such as $q=3$ and $4$ for Potts model) which would account for
4NN being weakly first order instead of second order. 
Furthermore, the critical exponents place 4NN squarely into
the Ising universality class, so that it would seem like 
a lot of coincidence for the Ising criticality 
to be only a crossover on the way to an underlying 
first order phase transition.  
We conjecture
that, in fact, 4NN is not the only continuous transition for higher
order exclusions. This conjecture is based on the observation
that as the exclusion region grows, the system approaches a
continuum limit --- lattice becomes irrelevant. 
For continuum 2d systems melting, however, is driven by 
the topological defects (dislocations) and is of second 
(actually infinite) order, belonging to the
Kosterlitz-Thouless universality 
class~\cite{KoTh73,Be71a,Be71b,Nelson78,NeHa79,Young79}.  
Clearly, this can not be reconciled
with the LL prediction of a first order transition
for all exclusions above 2NN.  Thus, there must be other lattice
gases with exclusions larger than 5NN which will also exhibit 
a continuous phase transition. 
The only way to fit this  
into LL theory is to relax the so called ``Lifshitz
rule'' which forbids the representations for which 
the antisymmetric part of their direct product 
contains a vector representation.  
This, however, would completely destroy the LL theory since the 
number of possible representations which can then be used to 
construct an invariant LGW Hamiltonian will become infinite.

It is very difficult to understand the mechanism which
changes a phase transition from first order, predicted by
a mean-field theory, to second order when the fluctuations are taken
into account. 
In two dimensions, however, this seems to be a very common occurrence 
as is demonstrated by the 
3 and 4 state Potts models 
which undergo continuous transitions in spite of
third order invariants present in their mean-field free energies.  
In this work we find that another 
mechanism is also possible --- the fluctuations can 
invalidate the Lifshitz condition, transforming
a first order phase transition into a fluctuation induced 
second order one.  In view of our conjecture that 1NN, 2NN and 4NN
are not the only lattice gases which undergo a continuous phase transition, 
it will be very curious to see 
what other higher order exclusions will also exhibit a continuous transition
and what 
distinguishes these systems from the ones which undergo a discontinuous
order-disorder phase transition.

\begin{acknowledgments}
We would like to acknowledge J.A. Plascak for many
interesting conversations and several helpful suggestions.
This work was supported in part by the Brazilian science agencies
CNPq, CAPES, and FAPERGS. HCMF acknowledges the Universitat de
Barcelona during his stay within the CAPES/MECD agreement. 
\end{acknowledgments}

\bibliographystyle{prsty}
\bibliography{heitor,mais}

\begin{thebibliography}{10}

\bibitem{Hill}
T. {Hill}, {\em Statistical Mechanics} (Mcgraw-Hill, New York, 1956).

\bibitem{RiSo03}
F. Ritort and P. Sollich, Adv. {P}hys. {\bf 52},  219  (2003).

\bibitem{GaFi65}
D. {Gaunt} and M. {Fisher}, J. Chem. Phys. {\bf 43},  2840  (1965).

\bibitem{Burley72}
D.~M. Burley,  in {\em Phase transitions and critical phenomena.}, edited by C.
  Domb and M.~S. Green (Academic Press, London-New York, 1972), Vol.~2,
  Chap.~9, pp.\ 329--374.

\bibitem{BeNi67}
A. {Bellemans} and R. {Nigam}, J. Chem. Phys. {\bf 46},  2922  (1967).

\bibitem{LaCu03jcp}
L. {Lafuente} and J. {Cuesta}, J. Chem. Phys. {\bf 119},  10832  (2003).

\bibitem{Frenkel}
D. {Frenkel} and B. {Smit}, {\em Understanding Molecular Simulation} (Academic
  Press, N. York, 2000).

\bibitem{FeSw88}
A. Ferrenberg and R. Swendsen, Phys. Rev. Lett. {\bf 61},  2635  (1988).

\bibitem{NeBa99}
M. Newman and G. Barkema, {\em {M}onte {C}arlo methods in statistical physics}
  (Oxford University Press, New York, USA, 1999).

\bibitem{Janke02}
W. Janke,  in {\em Computer Simulations of Surfaces and Interfaces}, Vol.~114
  of {\em Proceedings of the NATO Advanced Study Institute}, edited by B.
  Dünweg, D.~P. Landau, and A.~I. Milchev (Kluwer, Dordrecht, 2003), pp.\
  137--157.

\bibitem{Fisher63}
M.~E. Fisher, J. Math. Phys. {\bf 4},  278  (1963).

\bibitem{BaEnTs80}
R.~J. Baxter, I.~G. Enting, and S.~K. Tsang, J. Stat. Phys. {\bf 22},  465
  (1980).

\bibitem{BaFi94}
A. Baram and M. Fixman, J. Chem. Phys. {\bf 101},  3172  (1994).

\bibitem{Temperley62}
H.~N.~V. Temperley, Proceedings of the Physical Society {\bf 80},  813  (1962).

\bibitem{Runnels65b}
L.~K. Runnels, Phys. Rev. Lett. {\bf 15},  581  (1965).

\bibitem{RuCo66}
L.~K. Runnels and L.~L. Combs, J. Chem. Phys. {\bf 45},  2482  (1966).

\bibitem{ReCh66}
F.~H. Ree and D.~A. Chesnut, J. Chem. Phys. {\bf 45},  3983  (1966).

\bibitem{BeNi66}
A. Bellemans and R.~K. Nigam, Phys. Rev. Lett. {\bf 16},  1038  (1966).

\bibitem{NiFa74a}
R.~M. Nisbet and I.~E. Farquhar, Physica {\bf 76},  259  (1974).

\bibitem{WoGo80}
D.~W. Wood and M. Goldfinch, J. Phys. A {\bf 13},  2781  (1980).

\bibitem{KiSc81}
W. {Kinzel} and M. {Schick}, Phys. Rev. B {\bf 24},  324  (1981).

\bibitem{PeSe88}
P.~A. Pearce and K.~A. Seaton, J. Stat. Phys. {\bf 53},  1061  (1988).

\bibitem{GuBl02}
W. Guo and H.~W.~J. Bl\"ote, Phys. Rev. E {\bf 66},  046140  (2002).

\bibitem{Racz80}
Z. R\`acz, Phys. Rev. B {\bf 21},  4012  (1980).

\bibitem{HuCh91}
C.-K. Hu and C.-N. Chen, Phys. Rev. B {\bf 43},  6184  (1991).

\bibitem{BiLa80}
K. {Binder} and D. {Landau}, Phys. Rev. B {\bf 21},  1941  (1980).

\bibitem{Meirovitch83}
H. Meirovitch, J. Stat. Phys. {\bf 30},  681  (1983).

\bibitem{HuMa89}
C.-K. Hu and K.-S. Mak, Phys. Rev. B {\bf 39},  2948  (1989).

\bibitem{Haas89}
T.~M. Haas, J. Stat. Phys. {\bf 54},  201  (1989).

\bibitem{Yamagata95}
A. Yamagata, Physica A {\bf 222},  119  (1995).

\bibitem{HeBl96}
J.~R. Heringa and H.~W.~J. Bl\"ote, Physica A {\bf 232},  369  (1996).

\bibitem{HeBl98}
J.~R. Heringa and H.~W.~J. Bl\"ote, Physica A {\bf 251},  224  (1998).

\bibitem{LiEv00}
D.-J. Liu and J.~W. Evans, Phys. Rev. B {\bf 62},  2134  (2000).

\bibitem{Burley61}
D.~M. Burley, Proc. Phys. Soc. (London) {\bf 77},  451  (1961).

\bibitem{Runnels67}
L.~K. Runnels, J. Math. Phys. {\bf 8},  2081  (1967).

\bibitem{Woodbury67}
G.~W. Woodbury~Jr., J. Chem. Phys. {\bf 47},  270  (1967).

\bibitem{RoVa91}
A. Robledo and C. Varea, J. Stat. Phys. {\bf 63},  1163  (1991).

\bibitem{HaWe05a}
H. Hansen-Goos and M. Weigt, J. Stat. Mech.  P04006  (2005).

\bibitem{LaCu03pre}
L. {Lafuente} and J. {Cuesta}, Phys. Rev. E {\bf 68},  066120  (2003).

\bibitem{Runnels72}
L.~K. Runnels,  in {\em Phase transitions and critical phenomena.}, edited by
  C. Domb and M.~S. Green (Academic Press, London-New York, 1972), Vol.~2,
  Chap.~8, pp.\ 305--328.

\bibitem{Baxter99}
R.~J. Baxter, Ann.Comb. {\bf 3},  191  (1999).

\bibitem{FeScEe05}
P. Fendley, K. Schoutens, and H. van Eerten, J. Phys. A: Math. Gen. {\bf 38},
  315  (2005).

\bibitem{Murch81}
G.~E. Murch, Philos. Mag. A {\bf 44},  699  (1981).

\bibitem{Chaturvedi86}
D.~K. Chaturvedi, Phys. Rev. B {\bf 34},  8080  (1986).

\bibitem{MeCaLoSc87}
P. Meakin, J.~L. Cardy, E. Loh, and D.~J. Scalapino, J. Chem. Phys. {\bf 86},
  2380  (1987).

\bibitem{ErFrJa88}
W. {Ertel}, K. {Frobose}, and J. {Jackle}, J. Chem. Phys. {\bf 88},  5027
  (1988).

\bibitem{JaFrKn91}
J. Jäckle, K. Froböse, and D. Knödler, J. Stat. Phys. {\bf 63},  249  (1991).

\bibitem{DiWaJe91}
R. Dickman, J.-S. Wang, and I. Jensen, J. Chem. Phys. {\bf 94},  8252  (1991).

\bibitem{WaNiPr93a}
J.-S. Wang, P. Nielaba, and V. Privman, Mod. Phys. Lett. B {\bf 7},  189
  (1993).

\bibitem{EiBa98}
E. Eisenberg and A. Baram, Europhys. Lett. {\bf 44},  168  (1998).

\bibitem{Dickman01}
R. Dickman, Phys. Rev. E {\bf 64},  016124  (2001).

\bibitem{HaWe05b}
H. Hansen-Goos and M. Weigt, J. Stat. Mech.  P08001  (2005).

\bibitem{PoDi05}
F.~Q. Potiguar and R. Dickman, cond-mat/0507114 (unpublished).

\bibitem{Gaunt67}
D.~S. Gaunt, J. Chem. Phys. {\bf 46},  3237  (1967).

\bibitem{Cowley79}
E. Cowley, J. Chem. Phys. {\bf 71},  458  (1979).

\bibitem{HuMa90}
C.-K. Hu and K.-S. Mak, Phys. Rev. B {\bf 42},  965  (1990).

\bibitem{HeBlLu00}
J.~R. Heringa, H.~W.~J. Bl\"ote, and E. Luijten, J. Phys. A: Math. Gen. {\bf
  33},  2929  (2000).

\bibitem{Panagiotopoulos05}
A.~Z. Panagiotopoulos, J. Chem. Phys. {\bf 123},  104504  (2005).

\bibitem{LandauBinder}
D. {Landau} and K. {Binder}, {\em A Guide to Monte Carlo Simulations in
  Statistical Physics} (Cambridge University Press, Cambridge UK, 2000).

\bibitem{Baker66}
B.~G. Baker, J. Chem. Phys. {\bf 45},  2694  (1966).

\bibitem{ReCh67}
F. {Ree} and D. {Chesnut}, Phys. Rev. Lett. {\bf 18},  5  (1967).

\bibitem{Mehta74}
M.~L. Mehta, J. Chem. Phys. {\bf 60},  2207  (1974).

\bibitem{NiFa74c}
R. Nisbet and I. Farquhar, Physica {\bf 73},    (1974).

\bibitem{DoScWaGr78}
E. {Domany}, M. {Schick}, J. {Walker}, and R. {Griffiths}, Phys. Rev. B {\bf
  18},  2209  (1978).

\bibitem{AkSh82}
E. {Aksenenko} and Y. {Shulepov}, J. Phys. A: Math. Gen. {\bf 15},  2515
  (1982).

\bibitem{Slotte83}
P.~A. Slotte, J. Phys. A {\bf 16},  2935  (1983).

\bibitem{Huckaby83}
D.~A. Huckaby, J. Stat. Phys. {\bf 33},  23  (1983).

\bibitem{AkSh84}
E. {Aksenenko} and Y. {Shulepov}, J. Phys. A: Math. Gen. {\bf 17},  2109
  (1984).

\bibitem{AmKaGu84}
J. Amar, K. Kaski, and J.~D. Gunton, Phys. Rev. B {\bf 29},  1462  (1984).

\bibitem{WaNiPr93b}
J.-S. Wang, P. Nielaba, and V. Privman, Physica A {\bf 199},  527  (1993).

\bibitem{Uebing94}
C. {Uebing}, Physica A {\bf 210},  205  (1994).

\bibitem{ScLaCu03}
M. {Schmidt}, L. {Lafuente}, and J. {Cuesta}, J. Phys.: Condens. Matter {\bf
  15},  4695  (2003).

\bibitem{BaEi84}
N. Bartelt and T. Einstein, Phys. Rev. B {\bf 30},  5339  (1984).

\bibitem{MuZi77}
E. Müller-Hartmann and J. Zittarz, Z. Phys. B {\bf 27},  261  (1977).

\bibitem{HePr74}
O.~J. Heilmann and E. Praestgaard, J. Phys. A: Math. Gen. {\bf 7},  1913
  (1974).

\bibitem{OrBe82}
J. Orban and D. van Belle, J. Phys. A {\bf 15},  L501  (1982).

\bibitem{BeOr66}
A. Bellemans and J. Orban, Phys. Rev. Lett. {\bf 17},  908  (1966).

\bibitem{NiFa74b}
R.~M. Nisbet and I.~E. Farquhar, Physica {\bf 76},  283  (1974).

\bibitem{EiBa05}
E. Eisenberg and A. Baram, Europhys. Lett. {\bf 71},  900  (2005).

\bibitem{HaSt73}
C.~K. Hall and G. Stell, Phys. Rev. A {\bf 7},  1679  (1973).

\bibitem{FiNi82}
M.~E. Fisher and A. {Nihat Berker}, Phys. Rev. B {\bf 26},  2507  (1982).

\bibitem{Baxter73}
R.~J. Baxter, Journal of Physics C: Solid State Physics {\bf 6},  L445  (1973).

\bibitem{MiSt74}
L. Mittag and M.~J. Stephen, Journal of Physics A: Mathematical, Nuclear and
  General {\bf 7},  L109  (1974).

\bibitem{KoTh73}
J.~M. Kosterlitz and D.~J. Thouless, Journal of Physics C: Solid State Physics
  {\bf 6},  1181  (1973).

\bibitem{Be71a}
V.~L. Berezinskii, Zh. Eksp. Teor. Fiz {\bf 61},  1144  (1971).

\bibitem{Be71b}
V.~L. Berezinskii, Sov. Phys. - JETP {\bf 34},  610  (1971).

\bibitem{LeDa90}
Y. Levin and K.~A. Dawson, Phys. Rev. A {\bf 42},  3507  (1990).

\bibitem{Le91}
Y. Levin, Phys. Rev. B {\bf 43},  10876  (1991).

\bibitem{JoKaKiNe77}
J.~V. Jos\'e, L.~P. Kadanoff, S. Kirkpatrick, and D.~R. Nelson, Phys. Rev. B
  {\bf 16},  1217  (1977).

\bibitem{Nelson78}
D.~R. Nelson, Phys. Rev. B {\bf 18},  2318  (1978).

\bibitem{NeHa79}
D.~R. Nelson and B.~I. Halperin, Phys. Rev. B {\bf 19},  2457  (1979).

\bibitem{Young79}
A.~P. Young, Phys. Rev. B {\bf 19},  1855  (1979).

\end{thebibliography}

\end{document}